# Is more Phase Segregation better for mixed halide perovskite devices: Spatial randomness, Ion migration, and Non-radiative recombination

*Abhimanyu Singareddy, Uday Kiran Reddy Sadula and Pradeep R. Nair*

Department of Electrical Engineering, Indian Institute of Technology Bombay, Powai, Mumbai, 400076, India

Email: abhir@ee.iitb.ac.in, email: prnair@ee.iitb.ac.in,

**ABSTRACT** Phase segregation is a critical phenomenon that influences the stability and performance of mixed halide perovskite based opto-electronic devices. In addition to the underlying physical mechanisms, the spatial pattern and randomness associated with the nanoscale morphology of phase segregation significantly influence performance degradation – a topic which, along with the multitude of parameter combinations, has remained too complex to address so far. Given this, with $MAPbI_{1.5}Br_{1.5}$ as a model system, here we address the influence of critical factors like the spatial randomness of phase segregation, influence of ion migration, and the effect of increased non radiative recombination at material interfaces. Interestingly, our analytical model and detailed statistical simulations indicate a unique trend – morphology evolution with increased phase segregation results, surprisingly, in a recovery in efficiency while non-radiative recombination at grain/domain boundaries results in efficiency degradation. Further, our quantitative and predictive estimates identify critical parameters for interface states beyond which device variability could be an important system level bottleneck. Indeed, these estimates are broadly applicable to systems which undergo phase segregation and have interesting implications to perovskite based optoelectronic devices – from stability concerns to engineering approaches that attempt to arrest phase segregation.

**TOC GRAPHICS**

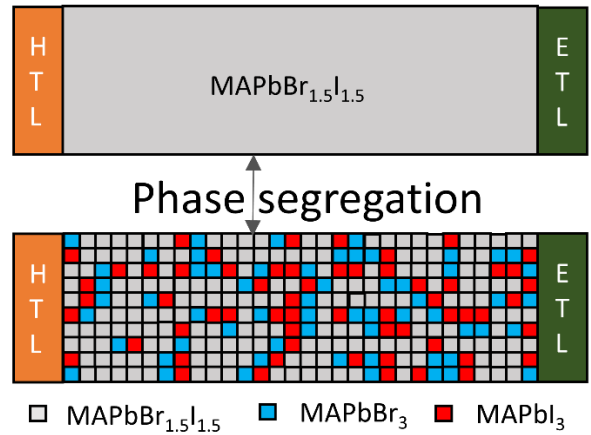

## I. INTRODUCTION

Perovskites has evoked tremendous research interest due to interesting applications in photovoltaics[1], memristors[2], photodetectors[3], Light Emitting Diodes[4], Tandem solar cells[5], etc. While bandgap tunability (through alloying different anions or cations or both)[6–10] is central to most such applications, phase segregation is a major concern. For example, the material $MAPbBr_xI_{1-x}$ is unstable under various conditions and phase segregates into iodide and bromide rich domains[11–13] (note that, as per convention[14,15], the term iodide represents $MAPbI_3$ (Methylammonium Lead Iodide) while bromide represents $MAPbBr_3$ (Methylammonium Lead Bromide) for the rest of this manuscript). This phenomenon is corroborated through optical measurements[11,12], where two distinct Photoluminescence (PL) peaks corresponding to iodide and bromide rich domains were observed. These two domains have different generation rates according to their absorption spectra. Literature also reports photo induced phase segregation[16], and electric bias induced phase segregation[17] in mixed perovskite structures. Interestingly, some reports indicate a reversal of phase segregation under dark conditions, although the time taken for recovery depends on light soaking conditions.[11,18,19] Given the prevalence of this phenomena and its broad appeal, a few questions are of natural interest to the community – (a) What causes phase segregation? (b) What is the eventual impact of phase segregation? (c) Is it possible to arrest phase segregation? Many reports tried to explain the cause of phase segregation (i.e., item a) by attributing it to the defects from halide ion vacancies which leads to the iodide and bromide rich segregated domains[20,21] and by using models based on polaron formation[22]. Alloying with a small amount of chlorine[23] and oxygen passivation[24] were reported to reduce the extent of phase segregation (i.e., item c). But the exact effect of phase segregation on performance metrics of a solar cell still remains elusive (i.e., item b) – even though it is the most critical system level concern. Indeed, there exist no previous studies in literature which explore the tradeoffs due to the nanoscale spatial randomness of phase segregation vis-à-vis lifetime degradation due to increased non-radiative recombination at interfaces of segregated domains along with ion migration effects. In this manuscript, we address this aspect through predictive models and detailed statistical simulations. Our results, surprisingly, indicate that the device performance, although severely affected initially, in fact improves with more phase segregation. As far as morphology or geometry of phase segregation is concerned, the more is indeed the better! – albeit with significant statistical fluctuations and dependence on material parameters. The most crucial among them being the increased non-radiative recombination at material or domain interfaces which causes significant degradation and variability in performance.

TOC figure (i.e., the figure provided with abstract) provides the schematic of pristine device structure (PIN solar cell)[25] which consist of Electron Transport Layer (ETL), Active layer, and Hole Transport Layer (HTL). ETL and HTL have their conventional definition and energy level alignments[26] (see Fig. 1a). For the pristine device, mixed halide perovskite ($MAPbBr_{1.5}I_{1.5}$) is the active layer. The figure provided in the abstract (and also Fig. 1b) is a representation of the device at some stage of phase segregation (with randomly chosen locations for bromide and iodide domains. These domains are of the size of tens of nm[18,27,28]). Here, for ease of analysis, we assume that the phase segregation can be treated as a transformation of a single material system (i.e., $MAPbBr_{1.5}I_{1.5}$) into a complex spatial pattern which involves three different materials (although, phase segregation need not result in generation of such pure domains)[12]. As such the material and electronic properties of the three different types of domains could be different (along with the ETL and HTL properties). Further, the presence of ion migration could significantly influence the electrostatics and hence the device performance. This could result in a plethora of parameter combinations and a detailed study could be a daunting task - not just to simulate but also to understand, assimilate, and convey the insights as well – precisely what we aim to achieve in this manuscript! The strategy is as follows: (a) we first explore the critical influence of spatial pattern or geometry of phase segregation (at low ion concentrations, see Fig. 1b). To this end, we assume that the three perovskites under consideration differ only in their band gaps (and hence photo-carrier generation rates) with corresponding energy level alignments shown in Fig. 1a. (b) Second, we consider the combined influence of ion migration and phase segregation (i.e., this is a case of high ion density, see Fig. 1c), and (c) finally, the critical effect of the combination of the phase segregation, high ion density, and the phase segregation dependent carrier lifetime degradation (due to a variety of factors which include increased non-radiative recombination at the phase segregated domains and/or at the interfaces between domains, see Fig. 1d). Such a hierarchical approach using models of increasing complexity indeed helps to quantitatively estimate the influence of phase segregation in mixed halide perovskite devices. Note that here we address the influence of abovementioned phenomena on the steady state efficiency (and not scan efficiency) of perovskite solar cells. It is widely accepted that hysteresis[29–31] is an interesting but transient phenomena (influenced by voltage scan properties), and is secondary in importance to steady state efficiency or stabilized efficiency[32,33] (where system is allowed to reach steady state at maximum power point conditions). In accordance with the above cited widely accepted literature, hysteresis is not addressed in this submission and could be taken up later.

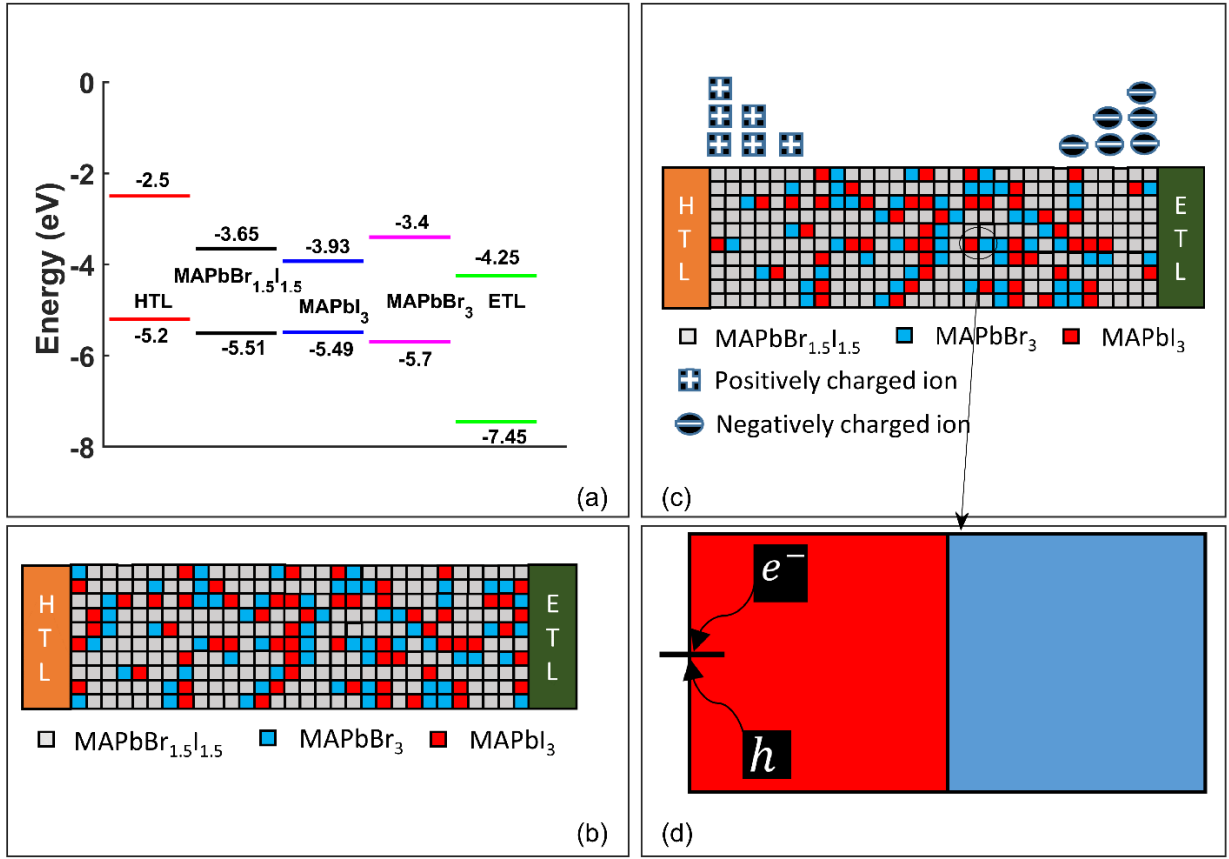

**Figure 1:** Model system (2D) to study the effect of phase segregation in mixed halide perovskite solar cells. (a) Energy levels of different materials[26]. The HTL properties correspond to Spiro-OMeTAD and ETL properties correspond to TiO₂. (b) Schematic of the model (with 300 domains in the active layer) used to study the effect of spatial randomness at nanoscale in the phase segregated device (low ion density case). (c) Scheme to investigate ion migration effects in a phase segregated device (high ion density case). (d) Schematic shows the increased non radiative recombination due to the traps present at the boundary of iodide domains along with the effect of ion migration in a phase segregated device.

## II. PREDICTIVE MODEL

Before we discuss the detailed simulation results, let us first attempt to arrive at analytical estimates for the performance metrics. Even with such a complex network of domains with heterojunctions, surprisingly, a few aspects related to phase segregation can be anticipated through insightful analysis of band level alignments and carrier transport. Note that the domains could be either iodide or bromide or mixed halide. As the energy level alignments indicate (see Fig. 1a), bromide domains are not energetically favorable for electrons. As a result, the electrons are expected to avoid the bromide domains and mostly flow through iodide or mixed halide domains. At low phase segregations, almost all paths from ETL could end up in HTL as the bromide domains could be a few and isolated. At very large phase segregations, the bromide domains could introduce significant impediment to photo-carrier collection. In addition, the efficiency of carrier collection is expected to depend on the internal electric field and hence the applied bias. Curiously, this could be significantly influenced by the presence of ions in the photo-active layer. Accordingly, any predictive model needs to account for the influence of carrier transport and ion influenced electrostatics in the perovskite active layer- as accomplished in this section.

(a)  **Hypothesis on $J_{SC}$**: Using parameters from Shockley-Queisser analysis[34], uniform optical absorption, and an optical loss[35] of 20%, we have $G_I = 4.53 \times 10^{21} \, cm^{-3} s^{-1}$, $G_M = 2.98 \times 10^{21} \, cm^{-3} s^{-1}$, and $G_B = 1.48 \times 10^{21} \, cm^{-3} s^{-1}$ as the photo-carrier generation rates for the iodide, mixed halide, and bromide domains, respectively. Let $N_I$, $N_M$, and $N_B$ denote the number of domains of iodide, mixed halide, and bromide, respectively. If we assume domains to have the same volume, then the level of phase segregation ($\vartheta$) can be defined as

$$\vartheta = \frac{N_I + N_B}{N_I + N_M + N_B}. \tag{1}$$

Assumption of volume conservation during phase segregation leads to $N_I + N_M + N_B = N_T$ where $N_T$ is a constant (300 in our simulations, see Fig. 1b) and $N_I = N_B$. The collection efficiency of photo-generated carriers depends on the density of ions present in the perovskite active layer. Accordingly, two limiting scenarios could be easily identified:

**(i) Low ion density**: Here, due to negligible amount of ionic charge in the perovskite active region, the device will behave like a classical PIN structure. The simulated E-B diagram for the pristine device (i.e., before any phase segregation happens) is shown as inset in Fig. 3a. Evidently, the dominant carrier collection mechanism is field assisted drift. For an assumed carrier mobility ($\mu$) of 2 $cm^2/Vs$, effective carrier lifetime ($\tau_{eff}$) of 100 ns, a built-in potential of 0.784 V, and active layer thickness of 300nm, the drift collection length ( $\mu\tau_{eff}E$, $E$ is the electric field) at short circuit conditions is ~52 μm. Since the drift collection length is much larger than the device thickness and neglecting the effect of potential barriers induced by bromide domains (see Fig. 1a) we have

$$J_{SC} = q \ \frac{(N_I G_I + N_M G_M + N_B G_B)}{N_T} \ l, \tag{2}$$

where $l$ is the thickness of active layer (300nm in our case). For the specific problem under consideration, we note that $G_M \sim (G_I + G_B)/2$. Accordingly, eq (2) indicates that

$$J_{SC} \sim q G_M l, \tag{3}$$

and hence the $J_{SC}$ will be independent of phase segregation.

**(ii) High Ion density**: In this case, the ions screen the $V_{bi}$ and the negligible electric field is present in the perovskite active layer even during short circuit conditions. This is evident in the E-B diagram of the pristine device (see Fig. 4a, simulation results with high ion density). Consequently, photo-carrier collection is dominated by diffusion and the corresponding collection length is given as $\sqrt{D\tau_{eff}}$. For similar set of parameters, the diffusion collection length is about 700nm, which is larger (but not significantly larger) than the active layer thickness of 300nm. Hence, we expect a weak dependence of photo-carrier collection and hence the $J_{SC}$ on phase-segregation.

The above arguments indicate that regardless of the presence of ions in the active layer, the $J_{SC}$ is expected to show only a very weak dependence on phase segregation. This is a consequence of the fact that both the drift as well as diffusion collection lengths under short circuit conditions are larger than the active layer thickness. On the other hand, if the lifetime of carriers reduces to 1ns, the drift and diffusion collection lengths reduce to 520nm and 70nm, respectively. Accordingly, our model predicts significant reduction in $J_{SC}$ in the presence of lifetime degradation and high ion densities. We remark that the model predictions on $J_{SC}$ also depends directly on the band gaps of various phases and could vary if one were to start with a different mixed halide perovskite other than the one under consideration (in such a case only $G_M$ changes with no change in $G_I$ or $G_B$).

   **(b) Predictions for $V_{oc}$:** The detailed balance $V_{oc}$ of an isolated intrinsic domain is given as $V_{OC} = \frac{2kT}{q} \ln\left(\frac{G\tau_{eff}}{n_i}\right)$ where $G$, $n_i$, and $\tau_{eff}$ are the generation rate, intrinsic concentration, and the effective carrier lifetime in the active layer, respectively. Accordingly, the detailed balance $V_{oc}$ of mixed halide, bromide, and iodide domains are $V_{OC,MX} = 1.36V$, $V_{OC,B} = 1.73V$, and $V_{OC,I} = 1.08V$, respectively. Naturally, the detailed balance $V_{oc}$ of a device without any phase segregation will be 1.36V. For 100% phase segregation, only iodide and bromide domains will be left as active components. Here, one may expect significant contribution by bromide domains to $V_{OC}$. However, such a strong influence of bromide domains is unlikely due to the fact that the iodide domains act as potential well for electrons (see energy level alignment in Fig. 1a). Accordingly, carrier accumulation is expected in iodide domains and the recombination in the same domains could dictate the $V_{oc}$. Applying the principle of detailed balance, we have

$$N_T G_M = \frac{N_I \Delta n}{\tau_{eff}}, \tag{4}$$

where the LHS denotes the net photo-carrier generation in the device while the RHS is the net carrier recombination in the iodide domains. Here $\Delta n$ denotes the excess carrier density and $\tau_{eff}$ denotes the effective carrier lifetime. As iodide domains act as potential wells for carriers, recombination in other domains is ignored in this analysis (validity of the same will be evident once we discuss numerical simulation results). Under such conditions, the $V_{OC}$ is nothing but the split in quasi-Fermi levels in the intrinsic iodide domains. Accordingly, for $\vartheta \neq 0$, Eqs. (1)-(4) lead us to

$$V_{OC}(\vartheta) = \frac{2kT}{q} ln\left(\frac{2G_M \tau_{eff}}{\vartheta\, n_{iI}}\right), \tag{5}$$

where $n_{iI}$ is the intrinsic concentration in the iodide domain. Eq. (5) predicts that the $V_{oc}$ varies as ln ($\vartheta$) and hence is not expected to be strongly dependent on phase segregation (except when $\vartheta$ is small). We note that the predicted $V_{oc}$ dependence on phase segregation in mixed halide perovskites (i.e., Eq. (5)) is distinctly different from the $V_{oc}$ scaling trends observed in multi-component organic solar cells[36,37] (where the influence of all components are explicit due to the inherent nature of non-geminate recombination at interfaces). Although only a weak dependence is predicted by eq. (5), some variations in $V_{oc}$ are expected due to the spatial pattern of phase segregation and the influence of band alignment of the active layer components with the contact layers (see Fig. 1a). Further, it is interesting to observe that the prediction for $V_{oc}$ (i.e., eq. (5)) is valid irrespective of the ion density in the active layer. This is due to the fact that under $V_{oc}$ conditions, regardless of the ion density, negligible electric field is present in the perovskite active layer and hence detailed balance analysis is expected to hold.

(c) **Hypothesis on Fill Factor (FF)**: For the PIN structures under consideration, the maximum power-point happens at conditions where the internal electric field could be very small. Accordingly, the carrier collection at such conditions will be dominated by diffusion. Depending on the extent of phase segregation, the carrier extraction from isolated iodide domains might be difficult at such conditions and hence this could lead to significant FF dependence on the spatial pattern of phase segregation. As the level of phase segregation increases, we expect more and more such isolated iodide domains which later merge. Large clusters and interconnected pathways (very similar to percolation phenomena[38]) could emerge for large phase-segregation. Under such conditions, we expect the FF to decrease first with phase segregation (with less statistical fluctuations) and then as the iodide domains form large clusters or pathways the FF is expected to increase (with large statistical fluctuations as some such pathways might lead to much better carrier collection than others). As with the $V_{oc}$, this prediction is expected to hold regardless of the density of ions in the active layer as carrier collection is significantly influenced by diffusion at $V_{mpp}$ (maximum power point) conditions – regardless of whether the ion density is significant or not.

(d) **Efficiency**: Summing up, in spite of the complexities associated with multitude of parameter combinations, we expect $J_{sc}$ to be fairly independent of phase segregation (for all cases other than simultaneous occurrence of large ion density and significant lifetime degradation). The $V_{oc}$ could show a decreasing, although not very dramatic, trend with phase segregation. The FF could first decrease and then increase with phase segregation. Overall, the efficiency is then expected to follow the trends of $J_{sc}$ and FF. It is expected to reach a minimum and then increase with further phase segregation – along with significant fluctuations. For the specific cases of lifetime degradation coupled with large ion densities, we expect a decrease in efficiency with phase-segregation. These predictions will be now compared against detailed numerical simulations.

## III. SIMULATIONS

Numerical simulation of a multi-heterojunction device with random domains is a computationally challenging job, and the following simulation methodology is adopted: The perovskite active layer is assumed to consist of several individual domains, where the material properties of each domain can be uniquely defined. Self-consistent numerical solution of Poisson and Carrier continuity equations[39] in two dimensions and steady state conditions is used to explore the effect of phase segregation with explicit consideration of heterojunctions between various domains. The presence of ions is also self-consistently accounted through the Poisson's equation. Here the 300nm thick perovskite layer (see Fig. 1b) is divided into $N_T = 300$ domains each of $10nm \times 10nm$ size[27,28]. Upon illumination, a certain fraction of the photo-active material MAPbBr$_{1.5}$I$_{1.5}$ gets phase segregated into MAPbI$_3$ and MAPbBr$_3$ domains. For ease of analysis, we consider only the case of complete phase segregation with an implicit assumption of volume conservation – i.e., two MAPbBr$_{1.5}$I$_{1.5}$ domains phase segregate into two domains - one MAPbI$_3$ and one MAPbBr$_3$ (in reality, the domains need not be purely iodide or bromide[11]. However, such a scenario is beyond the scope of this submission and could be handled with an effective set of material parameters). For a given volume fraction of such phase segregation, first the corresponding bromide domains are placed randomly, then iodide domains are identified within the neighborhood of respective bromide domains. To gain insights into the statistics and influence of spatial pattern of Br/I domains on solar cell performance, different random manifestations were simulated (20 different spatial distributions) for each volume fraction of phase segregation.

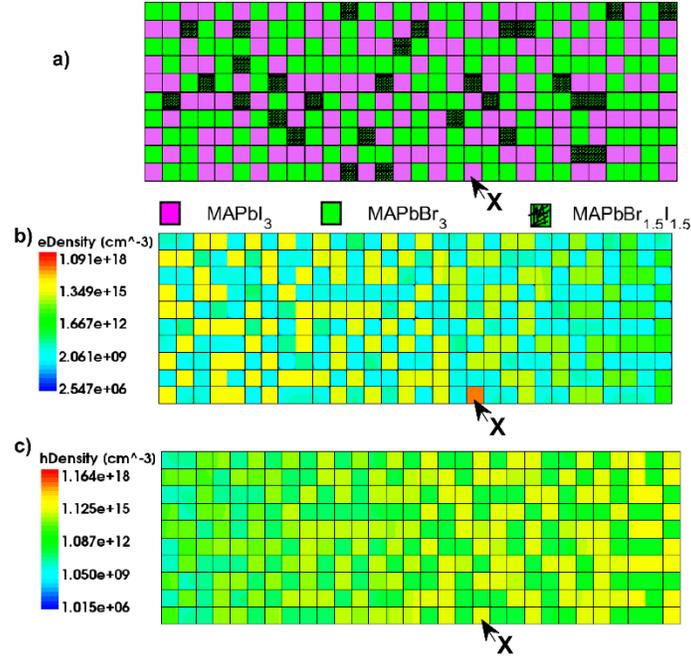

**Figure 2:** Dependence of carrier densities on the spatial configuration of phase segregation. a) Schematic of a configuration with 90% phase segregation to show the placement of different domains in the system (only the active layer is shown). (b) Electron and (c) hole density profiles plotted for (a) at maximum power point voltage. Clearly, iodide domains acts as potential wells where the carriers accumulate -in accordance with the energy level diagram shown in Fig. 1a. This is particularly evident for the domain indicated as X which is surrounded by bromide domains.

Figure 2a shows spatial configuration of a device at 90% phase segregation. Also shown are the electron and hole carrier densities. Here, we find that the carrier densities are more in the iodide domains – in tune with the energy level diagram shown in Fig. 1a and as predicted in Section II. Accordingly, it is evident that carrier transport will be predominantly through iodide domains and hence interconnectivity of such domains and carrier recombination at domain interfaces are expected to play a significant role in the device performance – which will be explored through models of hierarchical complexity in the following sections.

As discussed in Section I, the combinations of material parameters and spatial patterns of phase-segregation is too large to attempt an exhaustive simulation study. Still, aided with the insights from analytical model (Section II), we address the following effects – (a) influence of nanoscale geometry or spatial pattern of phase segregation, (b) Combined effect of ion migration and phase segregation, and (c) critical influence of lifetime degradation in conjunction with ion migration. Further, as mentioned before, transient effects like hysteresis are secondary to steady state efficiency and hence is reserved for later studies. As a calibration, we first performed simulations of PIN structures with only one active material. These characteristics are as expected, and the $V_{oc}$ values compare very well with the analytical predictions. Below, we discuss simulation results related to phase segregation:

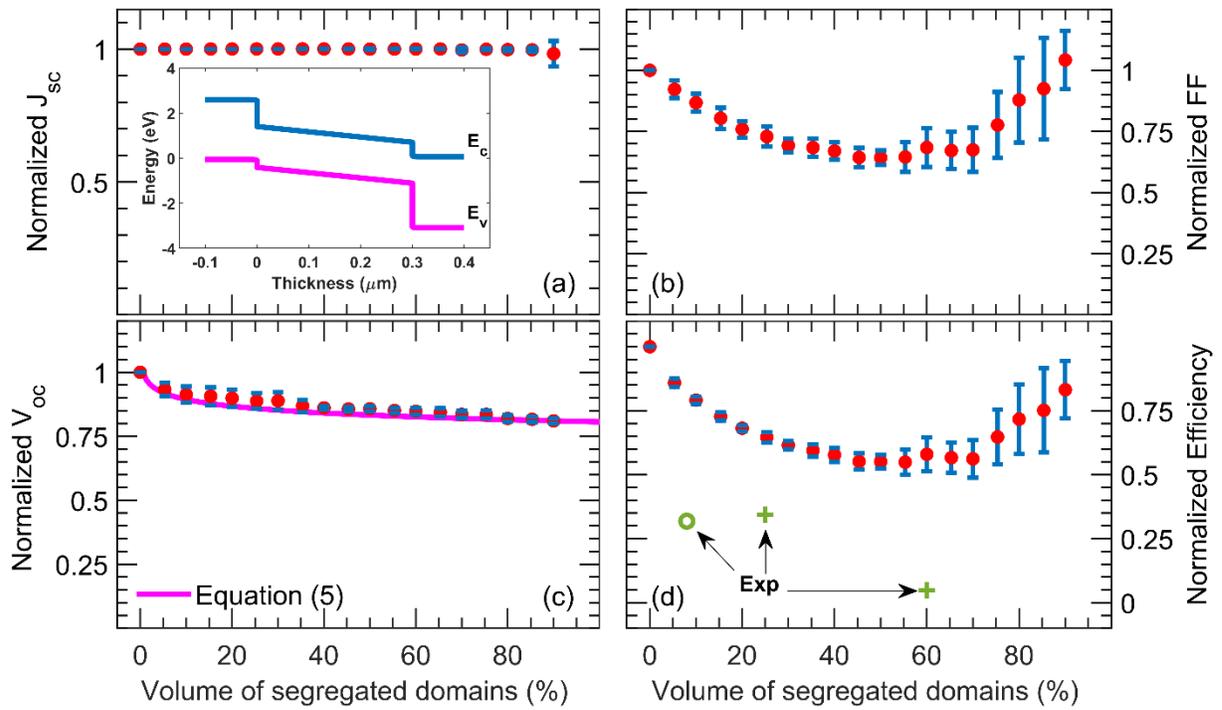

**Figure 3.** Effect of spatial randomness of nanoscale morphology associated with phase segregation on the performance parameters of a solar cell. (a) Short circuit current ($J_{sc}$, normalized to 14.31 $mA/cm^2$), Inset figure represents the equilibrium band diagram of the mixed halide perovskite device before segregation, (b) Open circuit voltage ($V_{oc}$, normalized to 1.356 V), (c) Fill Factor (FF, normalized to 61.21), and (d) efficiency (normalized to 11.88%). Black arrows highlight experimental data from literature: green circle symbol denotes data from ref 57, and green plus symbols denote data from ref 47. Note that all plots show the mean and standard deviation of the parameters.

**(a) Geometry of Nanoscale Phase segregation**: We first explore the influence of the spatial pattern of nanoscale phase segregation on the solar cell performance. The statistics of such simulations for various levels of phase segregation is provided in Fig. 3. For each volume fraction of phase segregation, we performed numerical simulations of 20 different spatial configurations and the mean and standard deviation of each data set is plotted. Interestingly, broad features of these results are in accordance with the analytical model, as listed below:

(a) The $J_{sc}$ is almost independent of phase segregation (shown in Fig. 3a) as the carrier collection length is much larger than device thickness, as predicted in the previous section. We note that reduction in $J_{sc}$ is seen in some configurations associated with large fraction of phase segregation. In such cases, we find that the bromide domains create significant obstruction to current collection.

(b) $V_{oc}$ trends (see Fig. 3c) are also reasonably well anticipated by the analytical model (eq. 5). This is not a surprise as eq. (5) was derived using the principal of detailed balance. Further, as expected, the fluctuations in $V_{oc}$ decrease as iodide domains increase in number.

(c) Our analytical model accurately predicts the FF trends (see Fig. 3b) – both the mean and the standard deviation. FF degrade initially and then improve with an increase in phase segregation. Further, we observe that for large phase segregation, the FF could be even better than that of the pristine device – which is a direct consequence of the influence of contact layer properties. For such cases, carrier collection is mainly through interconnected iodide domains in direct contact with transport layers. PIN schemes with iodide perovskite as the only active layer component results in much better FF as compared to devices with either bromide or mixed halides.

(d) Part (d) of Fig. 3 indicates that phase-segregation leads to an initial degradation of efficiency and then a recovery with significant variations – which was also predicted by our analytical model. For low phase segregation volumes, the iodide clusters are still isolated which results in increased recombination and hence decrease in efficiency. The performance recovers significantly for larger volume fractions of phase segregation. This improvement is contributed mainly by the emergence of carrier collection pathways where iodide domains are interconnected which results in better carrier collection and hence an increase in FF.

(e) Even though the performance recovers for larger values of phase segregation, the associated variation also becomes quite significant. This indicates that there could be significant device-device variability for larger phase segregations.

(f) Finally, the available experimental data in terms of efficiency degradation and phase segregation is plotted in Fig. 3d. The reported efficiency degradation is much worse than the predictions from numerical simulations – even after accounting for the fluctuations. This indicates that beyond spatial randomness, other phenomena could influence efficiency degradation associated with phase segregation. Nevertheless, the results shown in Fig. 3d indicates a limiting case – in the absence of any material and interface degradation, efficiency recovers with phase segregation.

**(b) Influence of Ion migration**: Perovskite materials are often plagued by ion migration which significantly affects the performance[40]. These ions under the influence of electric field could accumulate at the interface with contact layers and hamper the performance of the device. This is due to the low activation energies[41] of the ions in the lattice and hence can be easily influenced by factors such as external bias, light, heat etc., The charged species i.e. the ions or the vacancies left by them could be either positive ($N_{I,p}$) or negatively ($N_{I,n}$) charged.[41,42] In order to study the effects of migration of ions in the phase segregated device, we consider ions (both positive and negative) in the perovskite active region (with an average ion density $N_I = 10^{18}\ cm^{-3}$). These ions re-distribute under the influence of $V_{bi}$ and accumulate at the respective interfaces thus forming an ionic dipole in the active region (see schematic of Fig. 4a), which results in a flat band condition in the bulk of active region (see Energy band diagram shown in Fig. 4a). Accordingly, the dominant carrier transport mechanism changes from drift to diffusion.

The ionic charge distribution in perovskite active layer varies as a function of applied bias – indeed, ionic charge manifests as an ionic dipole in the active layer whose electric field opposes that of the built-in electric field, thus reducing the net electric field. To illustrate the same, the ionic dipole charge is plotted in Fig. 4b. Here the term dipole charge denotes the net ionic charge over half the width of perovskite active layer – i.e., the ionic dipole charge is nothing but $\int_0^{l/2} \rho_I(x)dx$, where $\rho_I(x) = N_{I,p}(x) - N_{I,n}(x)$ and $l$ is the thickness of the active layer (note that the ionic charges $N_{I,p}(x), N_{I,n}(x)$ are both position and bias dependent.[29,43] As expected, the ionic dipole charge vanishes at $V = V_{bi} = 0.8\ eV$ (see Fig. 4b). For $V < V_{bi}$, the negative ionic charges accumulate near ETL while positive ionic charge accumulates near HTL. Hence, the ionic dipole is positive; while the polarity reverses for $V > V_{bi}$ as negative ions now accumulate near HTL (and positive ions accumulate near ETL).

As mentioned before, the main focus of this manuscript is to explore the influence of phase segregation in the presence of ion migration and increased non-radiative recombination at domain interfaces. The corresponding numerical simulation results are provided in Fig. 4c, d where part

(c) correspond to the case of low ion density ($N_I = 0$), while part (d) correspond to the case of high ion density ($N_I = 10^{18} cm^{-3}$). Note that only the variation of efficiency is plotted for both scenarios.

We first focus on the results with $S = 0$ shown in Figs. 4c, d. A quick comparison indicates that the efficiency variation with phase segregation, in the absence of increased non-radiative recombination at domain interfaces (i.e., $S = 0$), shows similar features – regardless of the ion density, and in accordance with our predictive model. Minor deviations are observed in $J_{sc}$ with respect to the device without ions (which is expected as the carrier collection is now diffusion dominated), while the $V_{oc}$ trends are almost same. We note that these trends are in accordance with the model predictions. Even in the presence of ions, for $S = 0$, we find that efficiency variation in a phase segregated device follows a universal trend – an initial decrease with small fluctuations followed by a recovery with large fluctuations (see the data set corresponding to $S = 0$ in Fig. 4d).

**(c) Influence of increased non-radiative recombination** - The results so far indicate that the efficiency could recover under large values of phase segregation. This is due to critical influence of iodide domains in carrier collection and hence the FF. However, literature suggests that the non-radiative recombination increases in a device after phase segregation.[44] It could be due to deep traps at the domain boundaries/interfaces and/or increased non-radiative recombination in the phase segregated domains. An important parameter in this regard is the carrier recombination velocity (S) at the interfaces. Semiconductor device theory[45] indicates that $S \sim c N_{IT}$, where $c$ is the carrier capture coefficient and $N_{IT}$ is the interface trap density (i.e., areal density). In this regard, a wide range of parameters are reported in literature.[46,47] For example, the typical carrier capture coefficients are in the range $10^{-7}$ to $10^{-10} cm^3/s$, and the bulk trap densities are of the order of $10^{14} - 10^{16} \ cm^{-3}$. During phase segregation, one may expect higher density of bulk traps near the interface- say, of the order of $10^{15} - 10^{17} \ cm^{-3}$. Assuming interfaces of $nm$ thickness, the above correspond to broad range for $S$ of the order of $10^{-2} - 10^3 \ cm/s$. Given this broad range for S due to increased non-radiative recombination at interfaces, here we quantify the effect of the same on efficiency degradation with phase segregation.

The simulation results in Figs. 4c, d explore the influence of S on the efficiency as a function of phase segregation. There are several interesting trends:

(i) An increase in S, results in efficiency degradation – regardless of ion density (compare the cases with $S = 0$ and $S = 25 cm/s$),

(ii) the efficiency degradation with $S$ is more in the presence of large ion density (compare the results with $S = 250 cm/s$ in part (c) and (d)),

(iii) efficiency recovery with phase segregation is severely limited with large interface recombination (see the trends for $S = 250 cm/s$ in both parts (c) and (d)).

(iv) Literature results on efficiency degradation are in broad agreement with our simulation results (see Fig. 4c, d; green symbols denote experimental data from literature). Indeed, this work, for the first time, quantitatively predicts the efficiency degradation due to phase segregation in mixed halide perovskite devices.

The trends in Figs. 4c, d can be explained as follows: As discussed in the previous section, presence of mobile ions at high density flattens the bands which makes diffusion a dominant mode of transport. Further, it is well known[48] that the effective carrier lifetime in a domain is given as $\tau_{eff} \sim d/4S$, where d is the size of domains. Hence, any reduction in the lifetime (say, due to non-radiative recombination at domain interfaces) will further reduce the overall diffusion length of carriers. For $S = 250 \ cm/s$ or an effective lifetime of $1ns$ in iodide domains (with $d = 10nm$), we have $l_{diff} = \sqrt{D\tau_{eff}}$ which is around $72 \ nm$. This length is much shorter than the thickness of the device. So, as the number of iodide domains increase, the carrier recombination increases which results in a drastic reduction in $J_{sc}$ with respect to the volume of segregated domains.

Interestingly, as $\tau_{eff} \sim d/4S$, even modest amounts of interface degradation results in significant reduction in effective carrier lifetime. For example, the carrier lifetime degrades by an order of magnitude (i.e., from 100ns to 10ns) even with $S = 25 \ cm/s$. As discussed before, with typical carrier capture coefficients in the range $10^{-7}$ to $10^{-10} cm^3/s$,[46,47] the above correspond to an interface state trap density ($N_{IT}$) of around $5 \times 10^8$ to $5 \times 10^{11} \ cm^{-2}$. This is a result of immense importance as there is a strong correlation between domain size, interface state density and hence the effective lifetime. As the domain sizes are small (of the order of $10 \ nm$)[18,27,28], such low values of interface state density (equivalently S) lead to low effective lifetimes[49,50]. Hence it is important to both arrest phase segregation and passivate the interfaces – which are essential towards long term stability of such devices.

Our results, indeed, provide several insightful quantitative predictions for efficiency loss as function of critical parameters like phase segregation, ion migration, and increased non-radiative recombination at interfaces. The trends in Figs. 4c, d indicate that even modest amounts of phase segregation (~10%) result in nearly 50% degradation in efficiency – regardless of the presence of ionic charges. Increased non-radiative recombination along with ion migration often result in about 75% reduction in the efficiency. Further, given the large range of reported capture coefficients, interface recombination is expected to introduce significant variability in mixed halide devices – regardless of the presence of mobile ions.

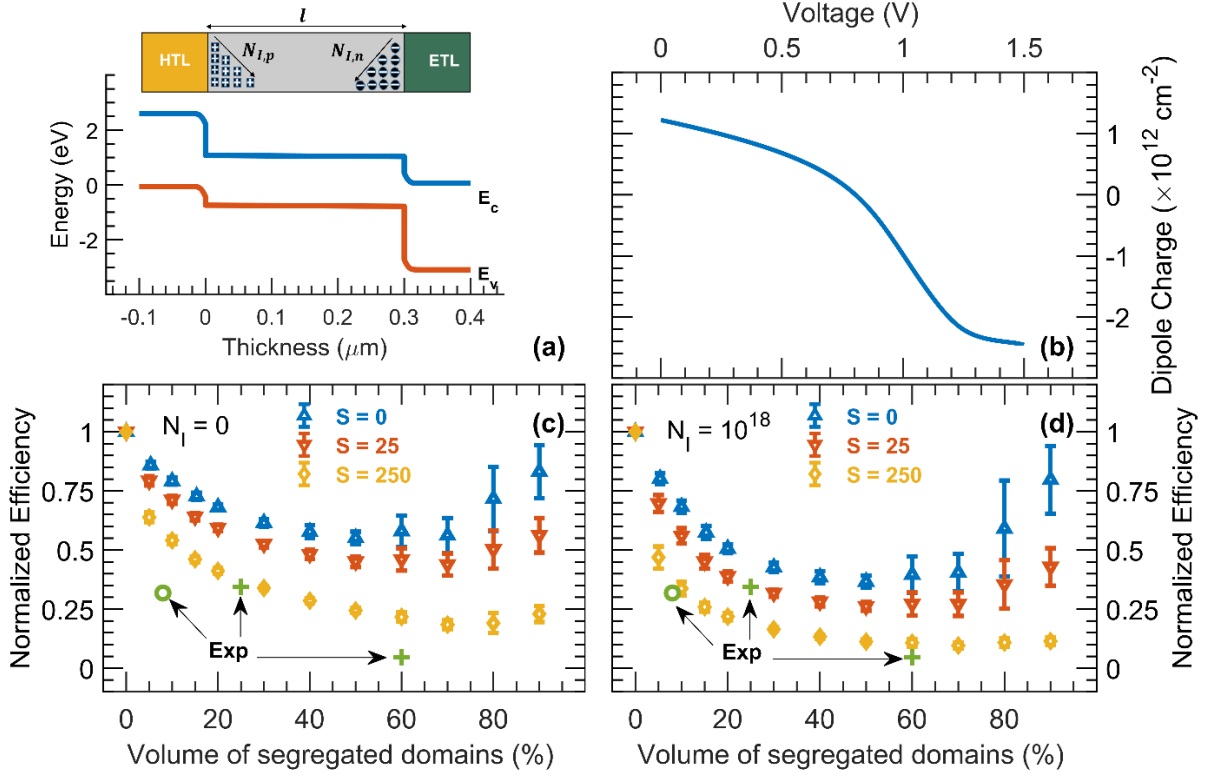

**Figure 4:** Influence of ions and the lifetime degradation in iodine domains on the efficiency of phase segregated device. (a) Schematic of the device with ions accumulated at the interfaces and the corresponding equilibrium band diagram (before segregation), (b) Variation of steady state ionic dipole charge at the interfaces as a function of applied bias, (c, d) Comparison of efficiency in low ($N_I = 0$, efficiency normalized to 11.88%) and high ion concentration ($N_I = 10^{18}\ cm^{-3}$, efficiency normalized to 12.47%) scenarios for three different recombination velocities (S) in a phase segregated device, i) $S = 0$, ii) $S = 25$, iii) $S = 250\,cm/s$. Black arrows refer to experimental data from literature: green circle symbol denotes data from ref 57, and green plus symbols denote data from ref 47. Note that c, d plots show the mean and standard deviation of the efficiencies. For each case and every volume fraction, at least 20 different random spatial configurations of phase segregation were simulated to obtain the statistics.

## IV. Discussions

Here we first compare the trends from simulations with available experimental data. Note that we addressed phase-segregation as transformation of a single component system into a three-component system. As such the conclusions should be broadly applicable to not just mixed halide perovskites but to other systems like quaternary system[16,17,51,52] which also exhibit phase segregation. While there are many reports on material characterization of systems undergoing phase segregation,[11,18,19,21,27,28,53] only a few studies have reported the effect of phase segregation on solar cells characteristics[17,52,54–56] – that too for individual devices and not any statistical study. Overall, the available experimental data can be summarized as: (i) fraction of phase segregation is estimated, without any reports on efficiency degradation,[11,19,23] (ii) efficiency degradation is reported without any information on fraction of phase segregation,[17,57] and (iii) both efficiency degradation and fraction of phase segregation is reported.[44,54] We note that our simulations are in good accordance with the data set (iii) – as shown in Fig. 4d. These results, for the first time, allows quantitative analysis of performance degradation vs important system level parameters like fraction of phase segregation and non-radiative recombination.

We remark that direct one-one quantitative comparisons with experimental results are not always possible due to multiple factors: (i) almost non-existent experimental results which explore the statistics of phase-segregation on efficiency degradation, and (ii) unknown/unspecified initial experimental initial conditions - in terms of measurement delays and the uncertainty associated with the extent of phase segregation (i.e., the initial measurement might correspond to an unknown $\vartheta \neq 0$). Still, isolated reports on degradation in $V_{oc}$ , FF and efficiency are available[13,17,54] in literature  which are in accordance with the results presented here. Observed degradation[54,56] in $J_{sc}$ could arise due to a degradation in effective carrier lifetime and/or due to ion migration, as explained in this manuscript. Curiously, an increase in $J_{sc}$ during phase segregation was reported recently[17] – which is not observed that regularly. Our analysis indicates that such a $J_{sc}$  increase, typically, would require formation of continuous paths of iodide from HTL to ETL at higher phase segregation volumes and/or a simultaneous improvement in the carrier lifetime as well. On the other hand, Mahesh et al.[51] reported through detailed balance analysis that the voltage loss ($V_{loss}$) by non-radiative recombination is dominant than $V_{loss}$ by phase segregation. But detailed balance analysis does not take into account the influence of spatial randomness and they could not explain the $J_{sc}$ degradation generally observed in the literature. Indeed, our model conclusively predicts not just the $V_{oc}$ loss due to the phase segregation and non-radiative recombination, but the associated statistical fluctuations as well. Further, our analysis is not limited to $V_{oc}$ alone, but provide predictive estimates for all performance parameters including the $J_{sc}$, FF, and efficiency.

In light of the above observations, our work provides insights for optimizing the performance of mixed halide perovskites. As mentioned before, detailed experimental studies to understand statistical effects of phase segregation are yet to reported. In this context, this work is very relevant as the influence of parametric variations is clearly elucidated in terms of the geometry or spatial configuration, influence of the ion migration, and the effectiveness of carrier collection due to increased non-radiative recombination at the interfaces of phase segregated domains. Evidently, the insights gained from the analytical model helped us to carefully select the parameter combinations for numerical simulations (results from ~1500 simulations were directly used in this submission) which allowed us to explore and understand the complexities associated with phase-segregation in a structured way. Further, available experimental data are in broad agreement with our simulations.

Having explored the key features of phase segregation, let us ponder about the broader implications. Our results indicate that phase segregation, ion migration and recombination at the domain interfaces have a critical influence towards the long-time operation and stability of perovskite-based devices. Our results indicate that the level of fluctuations is much smaller for lower values of phase segregation and hence it is always better to identify schemes[58] which can arrest phase segregation. At the low levels, the effect of ions has a minimal effect, as evident from the little variation seen in Fig. 4d. However, even at lower phase segregation levels, we need to be cautious about the degradation in effective lifetime due to the trap states at the interface of nanoscale domains. Indeed, our previous discussion indicates that S could vary over a broad range of $10^{-2} - 10^3 \ cm/s$. The results shown in Figs. 4c, d indicate that under such conditions, we expect an efficiency variation of the order of 50% over a broad range of phase segregation. All these indicate that the optimization of mixed halide perovskite devices or for that matter any system which undergoes segregation into domains is a challenging task. The nanoscale size of the domains, broad range in the capture coefficients and wide range of reported trap densities indicate that segregation induced variability could be an important optimization challenge. Hence to maximize performance, strategies should attempt to arrest phase segregation and to passivate any interface traps – else, along with the efficiency degradation, device-device variability could impede commercialization prospects of this technology. These observations are in agreement with the results in the literature.[51,57] On a related note, how would the design of tandem cells with mixed halides as top cells is going to be affected? Evidently, this is a complex task as the band gap and hence region of absorbed solar spectra could vary significantly that could render initial designs and optimizations ill suited. In this scheme also, arresting phase segregation has immense significance. Further, what might be the trends for phase-segregation with other mixed halides, say with $MAPbI_xBr_{3-x}$, where $x \neq 1.5$? For such cases, the broad trends as discussed here should hold. We note that $MAPbI_{1.5}Br_{1.5}$ has a band gap which resulted in net photo-carrier generation being a constant regardless of phase segregation. Although the same need not hold for other mole fractions, our analysis and simulation methodology could be easily extended to such cases (i.e., with mole fraction $x \neq 1.5$).

Finally, we wish to make a passing reference on similarity between carrier transport in phase segregation with the percolation concepts. In fact, the collection of photo-generated carriers at any level of phase segregation through interconnected nanoscale domains of iodides is very similar to percolation[38], except that classical percolation involves infinite potential barriers as against the finite potential barriers we encounter in halide phase segregation. Indeed, the fluctuations observed in this work are related to collection pathways and specific geometric pattern of various domains. Evidences of the influence of such percolative transport is observed in our simulations – especially in those related to the role of band level alignments. However, those results follow the broad trends discussed in this manuscript and will be communicated elsewhere. As mentioned before, hysteresis, a transient but important phenomen[31], is secondary in importance to steady state efficiencies at maximum power point conditions and, hence was not addressed in this submission. Future work could address related aspects and more complex 3D simulations.

## V. Conclusions

To summarize, here we provided quantitative estimates for the effect of phase segregation on mixed halide perovskite solar cells. Through a combination of analytical modeling and detailed numerical statistical simulations, we addressed the influence of critical aspects like spatial randomness of nanostructured heterojunctions, ion migration, and carrier collection efficiency. Our results indicate that while small amounts of phase segregation are indeed detrimental, performance recovers for larger values of phase segregation – a universal trend that holds over a broad range of material/electrical parameters, except when limited by non-radiative recombination at domain interfaces which could introduce significant device-device variability – a major system level concern from commercial perspectives. These predictions are well supported by experiments reported in literature. Our results are of immense significance for high efficiency tandem solar cells as phase segregation could be induced by illumination and could help in the design of optimal light soaking conditions which could result in stabilized efficiency for mixed halide perovskite solar cells. Indeed, the analytical predictions, simulation methodology and insights shared in this manuscript are broadly applicable to other perovskite systems where phase segregation could be a concern. Evidently, this work could help in the design of experiments to explore the statistics of phase segregation and interface engineering to reduce the influence of the same.

ASSOCIATED CONTENT

## AUTHOR INFORMATION

Email: abhir@ee.iitb.ac.in, email: prnair@ee.iitb.ac.in,

### ORCID

Abhimanyu singareddy: 0000-0002-9408-8386

Uday kiran Reddy Sadula: 0000-0003-4473-7402

Pradeep R. Nair:0000-0001-9977-2737

**Notes**

The authors declare no competing financial interest.

## ACKNOWLEDGMENT

This work was supported in part by Science and Engineering Research Board (SERB, project code: CRG/2019/003163), Department of Science and Technology, India. The authors would like to acknowledge CEN and NCPRE, IIT Bombay, India for computational facilities. We acknowledge the financial support of University Grants Commission (UGC), India. PRN also acknowledges Visvesvaraya Young Faculty Fellowship.

**REFERENCES**

(1)     Jeon, N. J.; Na, H.; Jung, E. H.; Yang, T. Y.; Lee, Y. G.; Kim, G.; Shin, H. W.; Il Seok, S.; Lee, J.; Seo, J. A Fluorene-Terminated Hole-Transporting Material for Highly Efficient and Stable Perovskite Solar Cells. *Nat. Energy* **2018**, *3* (8), 682–689. https://doi.org/10.1038/s41560-018-0200-6.

(2)     Xiao, X.; Hu, J.; Tang, S.; Yan, K.; Gao, B.; Chen, H.; Zou, D. Recent Advances in Halide Perovskite Memristors: Materials, Structures, Mechanisms, and Applications. *Adv. Mater. Technol.* **2020**, *5* (6), 1–29. https://doi.org/10.1002/admt.201900914.

(3)     Miao, J.; Zhang, F. Recent Progress on Highly Sensitive Perovskite Photodetectors. *J. Mater. Chem. C* **2019**, *7* (7), 1741–1791. https://doi.org/10.1039/C8TC06089D.

(4)     Zhang, Q.; Zhang, D.; Gu, L.; Tsui, K.-H.; Poddar, S.; Fu, Y.; Shu, L.; Fan, Z. Three-Dimensional Perovskite Nanophotonic Wire Array-Based Light-Emitting Diodes with Significantly Improved Efficiency and Stability. *ACS Nano* **2020**. https://doi.org/10.1021/acsnano.9b06663.

(5)     Bush, K. A.; Palmstrom, A. F.; Yu, Z. J.; Boccard, M.; Cheacharoen, R.; Mailoa, J. P.; McMeekin, D. P.; Hoye, R. L. Z.; Bailie, C. D.; Leijtens, T.; Peters, I. M.; Minichetti, M. C.; Rolston, N.; Prasanna, R.; Sofia, S.; Harwood, D.; Ma, W.; Moghadam, F.; Snaith, H. J.; Buonassisi, T.; Holman, Z. C.; Bent, S. F.; McGehee, M. D. 23.6%-Efficient Monolithic Perovskite/Silicon Tandem Solar Cells With Improved Stability. *Nat. Energy* **2017**, *2* (4),

1–7. https://doi.org/10.1038/nenergy.2017.9.

(6)  McMeekin, D. P.; Sadoughi, G.; Rehman, W.; Eperon, G. E.; Saliba, M.; Hörantner, M. T.; Haghighirad, A.; Sakai, N.; Korte, L.; Rech, B.; Johnston, M. B.; Herz, L. M.; Snaith, H. J. A Mixed-Cation Lead Mixed-Halide Perovskite Absorber for Tandem Solar Cells. *Science (80-. ).* **2016**, *351* (6269), 151–155. https://doi.org/10.1126/science.aad5845.

(7)  Rosales, B. A.; Hanrahan, M. P.; Boote, B. W.; Rossini, A. J.; Smith, E. A.; Vela, J. Lead Halide Perovskites: Challenges and Opportunities in Advanced Synthesis and Spectroscopy. *ACS Energy Lett.* **2017**, *2* (4), 906–914. https://doi.org/10.1021/acsenergylett.6b00674.

(8)  Noh, J. H.; Im, S. H.; Heo, J. H.; Mandal, T. N.; Seok, S. Il. Chemical Management for Colorful, Efficient, and Stable Inorganic-Organic Hybrid Nanostructured Solar Cells. *Nano Lett.* **2013**, *13* (4), 1764–1769. https://doi.org/10.1021/nl400349b.

(9)  Yi, C.; Luo, J.; Meloni, S.; Boziki, A.; Ashari-Astani, N.; Grätzel, C.; Zakeeruddin, S. M.; Röthlisberger, U.; Grätzel, M. Entropic Stabilization of Mixed A-Cation ABX3 Metal Halide Perovskites for High Performance Perovskite Solar Cells. *Energy Environ. Sci.* **2016**, *9* (2), 656–662. https://doi.org/10.1039/c5ee03255e.

(10) Suarez, B.; Gonzalez-Pedro, V.; Ripolles, T. S.; Sanchez, R. S.; Otero, L.; Mora-Sero, I. Recombination Study of Combined Halides (Cl, Br, I) Perovskite Solar Cells. *J. Phys. Chem. Lett.* **2014**, *5* (10), 1628–1635. https://doi.org/10.1021/jz5006797.

(11) Hoke, E. T.; Slotcavage, D. J.; Dohner, E. R.; Bowring, A. R.; Karunadasa, H. I.; McGehee, M. D. Reversible Photo-Induced Trap Formation in Mixed-Halide Hybrid Perovskites for Photovoltaics. *Chem. Sci.* **2015**, *6* (1), 613–617. https://doi.org/10.1039/c4sc03141e.

(12) Slotcavage, D. J.; Karunadasa, H. I.; McGehee, M. D. Light-Induced Phase Segregation in

Halide-Perovskite Absorbers. *ACS Energy Lett.* **2016**, *1* (6), 1199–1205. https://doi.org/10.1021/acsenergylett.6b00495.

(13) Braly, I. L.; Stoddard, R. J.; Rajagopal, A.; Uhl, A. R.; Katahara, J. K.; Jen, A. K. Y.; Hillhouse, H. W. Current-Induced Phase Segregation in Mixed Halide Hybrid Perovskites and Its Impact on Two-Terminal Tandem Solar Cell Design. *ACS Energy Lett.* **2017**, *2* (8), 1841–1847. https://doi.org/10.1021/acsenergylett.7b00525.

(14) Zhai, Y.; Wang, K.; Zhang, F.; Xiao, C.; Rose, A. H.; Zhu, K.; Beard, M. C. Individual Electron and Hole Mobilities in Lead-Halide Perovskites Revealed by Noncontact Methods. *ACS Energy Lett.* **2020**, *5* (1), 47–55. https://doi.org/10.1021/acsenergylett.9b02310.

(15) Chen, F.; Zhu, C.; Xu, C.; Fan, P.; Qin, F.; Gowri Manohari, A.; Lu, J.; Shi, Z.; Xu, Q.; Pan, A. Crystal Structure and Electron Transition Underlying Photoluminescence of Methylammonium Lead Bromide Perovskites. *J. Mater. Chem. C* **2017**, *5* (31), 7739–7745. https://doi.org/10.1039/c7tc01945a.

(16) Beal, R. E.; Hagström, N. Z.; Barrier, J.; Gold-Parker, A.; Prasanna, R.; Bush, K. A.; Passarello, D.; Schelhas, L. T.; Brüning, K.; Tassone, C. J.; Steinrück, H. G.; McGehee, M. D.; Toney, M. F.; Nogueira, A. F. Structural Origins of Light-Induced Phase Segregation in Organic-Inorganic Halide Perovskite Photovoltaic Materials. *Matter* **2020**, *2* (1), 207–219. https://doi.org/10.1016/j.matt.2019.11.001.

(17) Duong, T.; Mulmudi, H. K.; Wu, Y.; Fu, X.; Shen, H.; Peng, J.; Wu, N.; Nguyen, H. T.; Macdonald, D.; Lockrey, M.; White, T. P.; Weber, K.; Catchpole, K. Light and Electrically Induced Phase Segregation and Its Impact on the Stability of Quadruple Cation High Bandgap Perovskite Solar Cells. *ACS Appl. Mater. Interfaces* **2017**, *9* (32), 26859–26866. https://doi.org/10.1021/acsami.7b06816.

(18)   Draguta, S.; Sharia, O.; Yoon, S. J.; Brennan, M. C.; Morozov, Y. V.; Manser, J. M.; Kamat, P. V.; Schneider, W. F.; Kuno, M. Rationalizing the Light-Induced Phase Separation of Mixed Halide Organic-Inorganic Perovskites. *Nat. Commun.* **2017**, *8* (1). https://doi.org/10.1038/s41467-017-00284-2.

(19)   Yoon, S. J.; Draguta, S.; Manser, J. S.; Sharia, O.; Schneider, W. F.; Kuno, M.; Kamat, P. V. Tracking Iodide and Bromide Ion Segregation in Mixed Halide Lead Perovskites during Photoirradiation. *ACS Energy Lett.* **2016**, *1* (1), 290–296. https://doi.org/10.1021/acsenergylett.6b00158.

(20)   Yoon, S. J.; Kuno, M.; Kamat, P. V. Shift Happens. How Halide Ion Defects Influence Photoinduced Segregation in Mixed Halide Perovskites. *ACS Energy Lett.* **2017**, *2* (7), 1507–1514. https://doi.org/10.1021/acsenergylett.7b00357.

(21)   Barker, A. J.; Sadhanala, A.; Deschler, F.; Gandini, M.; Senanayak, S. P.; Pearce, P. M.; Mosconi, E.; Pearson, A. J.; Wu, Y.; Srimath Kandada, A. R.; Leijtens, T.; De Angelis, F.; Dutton, S. E.; Petrozza, A.; Friend, R. H. Defect-Assisted Photoinduced Halide Segregation in Mixed-Halide Perovskite Thin Films. *ACS Energy Lett.* **2017**, *2* (6), 1416–1424. https://doi.org/10.1021/acsenergylett.7b00282.

(22)   Mao, W.; Hall, C. R.; Bernardi, S.; Cheng, Y. B.; Widmer-Cooper, A.; Smith, T. A.; Bach, U. Light-Induced Reversal of Ion Segregation in Mixed-Halide Perovskites. *Nat. Mater.* **2021**, *20* (1), 55–61. https://doi.org/10.1038/s41563-020-00826-y.

(23)   Cho, J.; Kamat, P. V. How Chloride Suppresses Photoinduced Phase Segregation in Mixed Halide Perovskites. *Chem. Mater.* **2020**, *32* (14), 6206–6212. https://doi.org/10.1021/acs.chemmater.0c02100.

(24)   Fan, W.; Shi, Y.; Shi, T.; Chu, S.; Chen, W.; Ighodalo, K. O.; Zhao, J.; Li, X.; Xiao, Z.

Suppression and Reversion of Light-Induced Phase Separation in Mixed-Halide Perovskites by Oxygen Passivation. *ACS Energy Lett.* **2019**, *4* (9), 2052–2058. https://doi.org/10.1021/acsenergylett.9b01383.

(25)  Nandal, V.; Nair, P. R. Predictive Modeling of Ion Migration Induced Degradation in Perovskite Solar Cells. *ACS Nano* **2017**, *11* (11), 11505–11512. https://doi.org/10.1021/acsnano.7b06294.

(26)  Jena, A. K.; Kulkarni, A.; Miyasaka, T. Halide Perovskite Photovoltaics: Background, Status, and Future Prospects. *Chem. Rev.* **2019**, *119* (5), 3036–3103. https://doi.org/10.1021/acs.chemrev.8b00539.

(27)  Bischak, C. G.; Hetherington, C. L.; Wu, H.; Aloni, S.; Ogletree, D. F.; Limmer, D. T.; Ginsberg, N. S. Origin of Reversible Photoinduced Phase Separation in Hybrid Perovskites. *Nano Lett.* **2017**, *17* (2), 1028–1033. https://doi.org/10.1021/acs.nanolett.6b04453.

(28)  Brennan, M. C.; Draguta, S.; Kamat, P. V.; Kuno, M. Light-Induced Anion Phase Segregation in Mixed Halide Perovskites. *ACS Energy Lett.* **2018**, *3* (1), 204–213. https://doi.org/10.1021/acsenergylett.7b01151.

(29)  Van Reenen, S.; Kemerink, M.; Snaith, H. J. Modeling Anomalous Hysteresis in Perovskite Solar Cells. *J. Phys. Chem. Lett.* **2015**, *6* (19), 3808–3814. https://doi.org/10.1021/acs.jpclett.5b01645.

(30)  Tress, W.; Marinova, N.; Moehl, T.; Zakeeruddin, S. M.; Nazeeruddin, M. K.; Grätzel, M. Understanding the Rate-Dependent J-V Hysteresis, Slow Time Component, and Aging in CH3NH3PbI3 Perovskite Solar Cells: The Role of a Compensated Electric Field. *Energy Environ. Sci.* **2015**, *8* (3), 995–1004. https://doi.org/10.1039/c4ee03664f.

(31)  Habisreutinger, S. N.; Noel, N. K.; Snaith, H. J. Hysteresis Index: A Figure without Merit

for Quantifying Hysteresis in Perovskite Solar Cells. *ACS Energy Lett.* **2018**, *3* (10), 2472–2476. https://doi.org/10.1021/acsenergylett.8b01627.

(32) Khenkin, M. V; Katz, E. A.; Abate, A.; Bardizza, G.; Berry, J. J.; Brabec, C.; Brunetti, F.; Bulović, V.; Burlingame, Q.; Carlo, A. Di; Matheron, M.; Mcgehee, M.; Meitzner, R.; Nazeeruddin, M. K. Consensus Statement for Stability Assessment and Reporting for Perovskite Photovoltaics Based on ISOS Procedures. *Nat. Energy* **2020**, *5* (January). https://doi.org/10.1038/s41560-019-0529-5.

(33) Li, N.; Niu, X.; Chen, Q.; Zhou, H. Towards Commercialization: The Operational Stability of Perovskite Solar Cells. *Chem. Soc. Rev.* **2020**, *49* (22), 8235–8286. https://doi.org/10.1039/d0cs00573h.

(34) Shockley, W.; Queisser, H. J. Detailed Balance Limit of Efficiency of P-n Junction Solar Cells. *J. Appl. Phys.* **1961**, *32* (3), 510–519. https://doi.org/10.1063/1.1736034.

(35) Agarwal, S.; Nair, P. R. Performance Loss Analysis and Design Space Optimization of Perovskite Solar Cells. *J. Appl. Phys.* **2018**, *124* (18). https://doi.org/10.1063/1.5047841.

(36) Kang, H.; Kim, K. H.; Kang, T. E.; Cho, C. H.; Park, S.; Yoon, S. C.; Kim, B. J. Effect of Fullerene Tris-Adducts on the Photovoltaic Performance of P3HT:Fullerene Ternary Blends. *ACS Appl. Mater. Interfaces* **2013**, *5* (10), 4401–4408. https://doi.org/10.1021/am400695e.

(37) Khlyabich, P. P.; Burkhart, B.; Thompson, B. C. Efficient Ternary Blend Bulk Heterojunction Solar Cells with Tunable Open-Circuit Voltage. *J. Am. Chem. Soc.* **2011**, *133* (37), 14534–14537. https://doi.org/10.1021/ja205977z.

(38) Stauffer, D.; Aharony, A. *Introduction To Percolation Theory*; Taylor & Francis: London, 2018. https://doi.org/10.1201/9781315274386.

(39)    Synopsys. Sentaurus Device Simulation Tool. 2011.

(40)    Lee, J. W.; Kim, S. G.; Yang, J. M.; Yang, Y.; Park, N. G. Verification and Mitigation of Ion Migration in Perovskite Solar Cells. *APL Mater.* **2019**, *7* (4). https://doi.org/10.1063/1.5085643.

(41)    Eames, C.; Frost, J. M.; Barnes, P. R. F.; O'Regan, B. C.; Walsh, A.; Islam, M. S. Ionic Transport in Hybrid Lead Iodide Perovskite Solar Cells. *Nat. Commun.* **2015**, *6* (May), 2– 9. https://doi.org/10.1038/ncomms8497.

(42)    Sherkar, T. S.; Momblona, C.; Gil-Escrig, L.; Ávila, J.; Sessolo, M.; Bolink, H. J.; Koster, L. J. A. Recombination in Perovskite Solar Cells: Significance of Grain Boundaries, Interface Traps, and Defect Ions. *ACS Energy Lett.* **2017**, *2* (5), 1214–1222. https://doi.org/10.1021/acsenergylett.7b00236.

(43)    Neukom, M. T.; Schiller, A.; Züfle, S.; Knapp, E.; Ávila, J.; Pérez-Del-Rey, D.; Dreessen, C.; Zanoni, K. P. S.; Sessolo, M.; Bolink, H. J.; Ruhstaller, B. Consistent Device Simulation Model Describing Perovskite Solar Cells in Steady-State, Transient, and Frequency Domain. *ACS Appl. Mater. Interfaces* **2019**, *11* (26), 23320–23328. https://doi.org/10.1021/acsami.9b04991.

(44)    Balakrishna, R. G.; Kobosko, S. M.; Kamat, P. V. Mixed Halide Perovskite Solar Cells. Consequence of Iodide Treatment on Phase Segregation Recovery. *ACS Energy Lett.* **2018**, *3* (9), 2267–2272. https://doi.org/10.1021/acsenergylett.8b01450.

(45)    Pierret, R. F. Advanced Semiconductor Fundamentals. *Book* **1987**, *121* (7), 221.

(46)    Shi, J.; Li, Y.; Li, Y.; Li, D.; Luo, Y.; Wu, H.; Meng, Q. From Ultrafast to Ultraslow: Charge-Carrier Dynamics of Perovskite Solar Cells. *Joule* **2018**, *2* (5), 879–901. https://doi.org/10.1016/j.joule.2018.04.010.

(47) Fu, X.; Weber, K. J.; White, T. P. Characterization of Trap States in Perovskite Films by Simultaneous Fitting of Steady-State and Transient Photoluminescence Measurements. *J. Appl. Phys.* **2018**, *124* (7). https://doi.org/10.1063/1.5029278.

(48) Yang, Y.; Yan, Y.; Yang, M.; Choi, S.; Zhu, K.; Luther, J. M.; Beard, M. C. Low Surface Recombination Velocity in Solution-Grown CH 3 NH 3 PbBr 3 Perovskite Single Crystal. *Nat. Commun.* **2015**, *6*. https://doi.org/10.1038/ncomms8961.

(49) Chu, Z.; Yang, M.; Schulz, P.; Wu, D.; Ma, X.; Seifert, E.; Sun, L.; Li, X.; Zhu, K.; Lai, K. Impact of Grain Boundaries on Efficiency and Stability of Organic-Inorganic Trihalide Perovskites. *Nat. Commun.* **2017**, *8* (1), 1–8. https://doi.org/10.1038/s41467-017-02331-4.

(50) Nandal, V.; Nair, P. R. Ion Induced Passivation of Grain Boundaries in Perovskite Solar Cells. *J. Appl. Phys.* **2019**, *125* (17). https://doi.org/10.1063/1.5082967.

(51) Mahesh, S.; Ball, J. M.; Oliver, R. D. J.; McMeekin, D. P.; Nayak, P. K.; Johnston, M. B.; Snaith, H. J. Revealing the Origin of Voltage Loss in Mixed-Halide Perovskite Solar Cells. *Energy Environ. Sci.* **2020**, *13* (1), 258–267. https://doi.org/10.1039/c9ee02162k.

(52) Yang, Z.; Rajagopal, A.; Jo, S. B.; Chueh, C. C.; Williams, S.; Huang, C. C.; Katahara, J. K.; Hillhouse, H. W.; Jen, A. K. Y. Stabilized Wide Bandgap Perovskite Solar Cells by Tin Substitution. *Nano Lett.* **2016**, *16* (12), 7739–7747. https://doi.org/10.1021/acs.nanolett.6b03857.

(53) Elmelund, T.; Seger, B.; Kuno, M.; Kamat, P. V. How Interplay between Photo and Thermal Activation Dictates Halide Ion Segregation in Mixed Halide Perovskites. *ACS Energy Lett.* **2020**, *5* (1), 56–63. https://doi.org/10.1021/acsenergylett.9b02265.

(54) Samu, G. F.; Janáky, C.; Kamat, P. V. A Victim of Halide Ion Segregation. How Light Soaking Affects Solar Cell Performance of Mixed Halide Lead Perovskites. *ACS Energy*

*Lett.* **2017**, *2* (8), 1860–1861. https://doi.org/10.1021/acsenergylett.7b00589.

(55) Yang, X.; Yan, X.; Wang, W.; Zhu, X.; Li, H.; Ma, W.; Sheng, C. X. Light Induced Metastable Modification of Optical Properties in CH3NH3PbI3-XBrx Perovskite Films: Two-Step Mechanism. *Org. Electron. physics, Mater. Appl.* **2016**, *34*, 79–83. https://doi.org/10.1016/j.orgel.2016.04.020.

(56) Hu, M.; Bi, C.; Yuan, Y.; Bai, Y.; Huang, J. Stabilized Wide Bandgap MAPbBrxI3-x Perovskite by Enhanced Grain Size and Improved Crystallinity. *Adv. Sci.* **2015**, *3* (6), 6–11. https://doi.org/10.1002/advs.201500301.

(57) Penã-Camargo, F.; Caprioglio, P.; Zu, F.; Gutierrez-Partida, E.; Wolff, C. M.; Brinkmann, K.; Albrecht, S.; Riedl, T.; Koch, N.; Neher, D.; Stolterfoht, M. Halide Segregation versus Interfacial Recombination in Bromide-Rich Wide-Gap Perovskite Solar Cells. *ACS Energy Lett.* **2020**, *5* (8), 2728–2736. https://doi.org/10.1021/acsenergylett.0c01104.

(58) Knight, A. J.; Herz, L. M. Preventing Phase Segregation in Mixed-Halide Perovskites: A Perspective. *Energy Environ. Sci.* **2020**, 2024–2046. https://doi.org/10.1039/d0ee00788a.

(59) Singareddy, A.; Kiran, U.; Sadula, R.; Nair, P. R. Phase Segregation in Mixed Halide Perovskites- Is More Better ? 1–28.

(60) Agarwal, S.; Seetharaman, M.; Kumawat, N. K.; Subbiah, A. S.; Sarkar, S. K.; Kabra, D.; Namboothiry, M. A. G.; Nair, P. R. On the Uniqueness of Ideality Factor and Voltage Exponent of Perovskite-Based Solar Cells. *J. Phys. Chem. Lett.* **2014**, *5* (23), 4115–4121. https://doi.org/10.1021/jz5021636.

(61) Agarwal, S.; Nair, P. R. Device Engineering of Perovskite Solar Cells to Achieve near Ideal Efficiency. *Appl. Phys. Lett.* **2015**, *107* (12). https://doi.org/10.1063/1.4931130.

(62) Knight, A. J.; Wright, A. D.; Patel, J. B.; McMeekin, D. P.; Snaith, H. J.; Johnston, M. B.;

Herz, L. M. Electronic Traps and Phase Segregation in Lead Mixed-Halide Perovskite. *ACS Energy Lett.* **2019**, *4* (1), 75–84. https://doi.org/10.1021/acsenergylett.8b02002.

(63)   Bi, D.; Tress, W.; Dar, M. I.; Gao, P.; Luo, J.; Renevier, C.; Schenk, K.; Abate, A.; Giordano, F.; Correa Baena, J. P.; Decoppet, J. D.; Zakeeruddin, S. M.; Nazeeruddin, M. K.; Grätzel, M.; Hagfeldt, A. Efficient Luminescent Solar Cells Based on Tailored Mixed-Cation Perovskites. *Sci. Adv.* **2016**, *2* (1). https://doi.org/10.1126/sciadv.1501170.

(64)   Shen, J. X.; Zhang, X.; Das, S.; Kioupakis, E.; Van de Walle, C. G. Unexpectedly Strong Auger Recombination in Halide Perovskites. *Adv. Energy Mater.* **2018**, *8* (30). https://doi.org/10.1002/aenm.201801027.